# Numerical investigation of correlation functions for the $U_qSU(2)$ invariant spin-$\frac{1}{2}$ Heisenberg chain

Peter F. Arndt and Thomas Heinzel

*Physikalisches Institut*
*Universität Bonn, Nussallee 12, 53115 Bonn, Germany*

**Abstract**

We consider the $U_qSU(2)$ invariant spin-$\frac{1}{2}$ $XXZ$ quantum spin chain at roots of unity $q = \exp(\frac{i\pi}{m+1})$, corresponding to different minimal models of conformal field theory. We conduct a numerical investigation of correlation functions of $U_qSU(2)$ scalar two-point operators in order to find which operators in the minimal models they correspond to. Using graphical representations of the Temperley–Lieb algebra we are able to deal with chains of up to 28 sites. Depending on $q$ the correlation functions show different characteristics and finite size behaviour. For $m = \frac{2}{3}$, which corresponds to the Lee–Yang edge singularity, we find the surface and bulk critical exponent $-\frac{1}{5}$. Together with the known result in the case $m = 3$ (Ising model) this indicates that in the continuum limit the two-point operators involve conformal fields of spin $\frac{m-1}{m+1}$. For other roots of unity $q$ the chains are too short to determine surface and bulk critical exponents.

## 1. Introduction

We consider two-point correlation functions for a class of one-dimensional quantum models on a chain of $N$ sites defined in terms of the hamiltonian [1, 2]

$$H = -\sum_{i=1}^{N-1} e_i \, . \tag{1.1}$$

Here the $e_i$, $i = 1, 2, \ldots, N-1$, are generators of a Temperley–Lieb algebra $T_N(q)$ defined by the relations [3]

$$e_i \, e_i = (q + q^{-1}) \, e_i \tag{1.2a}$$

$$e_i \, e_{i\pm1} \, e_i = e_i \tag{1.2b}$$

$$e_i \, e_j = e_j \, e_i \qquad\qquad (j \neq i \pm 1) \, . \tag{1.2c}$$

In this article we numerically compute correlation functions of two-point operators that are elements of this algebra. Representing the generators by

$$e_i = -\frac{1}{2}\Big(\sigma_i^x \sigma_{i+1}^x \; + \; \sigma_i^y \sigma_{i+1}^y \; + \; \frac{q+q^{-1}}{2}\,(\sigma_i^z \sigma_{i+1}^z - 1) \; + \; \frac{q-q^{-1}}{2}\,(\sigma_i^z - \sigma_{i+1}^z)\Big) \tag{1.3}$$

these two-point operators become $U_qSU(2)$ invariant generalizations of the scalar operators $\vec{\sigma}_k \vec{\sigma}_l$. Here $\sigma_i^x$, $\sigma_i^y$, and $\sigma_i^z$ are Pauli matrices acting on site $i$. In the representation (1.3) $H$ turns into a $U_qSU(2)$ symmetric spin-$\frac{1}{2}$ $XXZ$ Heisenberg hamiltonian. Throughout this article $N$ is even to ensure that the ground state is unique.

If $q$ is real, the model is in a massive phase and in the limit $N \to \infty$ correlation functions of local operators can be computed [4]. For the $SU(2)$ symmetric model ($q = 1$) the correlation function of the operator $\sigma_k^z \sigma_l^z$ has been computed numerically using exact diagonalization [5, 6, 7] or Bethe ansatz [8]. The same correlation function has been calculated analytically in [9] for the periodic $XXZ$ chain (where the $U_qSU(2)$ symmetry is lost). For this case numerical results can be found in [10].

In this work we consider the hamiltonian $H$ and the two-point scalar operators $g_{k,l}$ (as defined in section 2) in terms of Temperley–Lieb generators. This includes the $U_qSU(2)$ symmetric case (1.3). Since in our calculations we use the Temperley–Lieb relations only, the results apply also to other representations besides this spin-$\frac{1}{2}$ chain [11]. We take $q = \exp(\frac{i\pi}{m+1})$ a root of unity. In this case the model is massless and the spectrum of $H$ contains the spectrum of hamiltonians corresponding to conformal field theories with central charge [12, 13, 14]

$$c = 1 - \frac{6}{m(m+1)} \, . \tag{1.4}$$

Although in the representation (1.3) the hamiltonian is not hermitian, its spectrum is real [1].

The two-point scalar operators are non-local and their continuum limit is not known for general $m$. However, for $m = 2$ and $m = 3$ the correlation functions have been derived analytically using quotients of the Temperley–Lieb algebra [2, 15]. In the former case the correlation functions vanish. In the latter case $H$ can be related to the hamiltonian of the Ising model and one finds two different non-vanishing correlation functions, whose continuum limit is given by correlation functions of operators with conformal dimensions $(h, \bar{h}) = (\frac{1}{2}, 0)$ and $(0, \frac{1}{2})$. For $m = 5$ there is a quotient of $T_N(q)$ which gives a three state Potts model, but the correlation function has not been computed in this case. In this article we present a numerical investigation of the correlation functions in the cases $m = 3, 5, \frac{2}{3}$ corresponding to Ising model, three state Potts model, and Lee–Yang edge singularity respectively. The first case is included to estimate the accuracy of the numerical results. For $m = 5$ we find four different correlation functions $\langle g_{k,l} \rangle$, depending on $k$ and $l$ odd or even. In the case $m = \frac{2}{3}$ there is only one correlation function. We attempt to find critical exponents and to identify conformal fields that correspond to the continuum limit of the two-point scalar operators. Due to a symmetry of the two-point scalar operator and from the known results in the Ising case one expects conformal fields with spin $\frac{m-1}{m+1}$. This can be confirmed in the case $m = \frac{2}{3}$.

For the computation of the correlation functions we use graphical representations of $T_N(q)$ on a path space and on boundary diagrams [16, 11]. In the representation on boundary diagrams one can easily restrict the configuration space to a space, which is related to the $U_qSU(2)$ scalars. For integer $m$ we can use a corresponding path representation space with a dimension that is further reduced. Whereas in the spin space approach (1.3) a reduction of the representation to the space of $U_qSU(2)$ scalars cannot be achieved easily. For the representation on boundary diagrams we can handle chains of up to 24 sites (independent of $m$), and up to 28 sites with the path space representation for $m = 5$. We extrapolate the data for $N \to \infty$ to correlations in the semi-infinite geometry, that is we keep one end of the chains fixed. From that we try to compute bulk and surface exponents.

The article organizes as follows. In section 2 we define the two-point scalar operator and review some results from [2]. The graphical representations of the Temperley–Lieb algebra we use are introduced in section 3. In section 4 we describe the calculation of ground state and correlation functions. We analyze the results of the computation for the cases $m = 3, 5, \frac{2}{3}$ in section 5, numerical data for the semi-infinite chain are listed in an appendix. Section 6 contains our conclusions.

## 2. Two-point scalar operators

Two-point $U_qSU(2)$ scalar operators have been derived in [2]. They can be defined recursively in terms of Temperley–Lieb generators $e_k$:

$$g_{k,k+1} = e_k - (q+q^{-1})^{-1} \qquad 1 \le k \le N-1 \qquad (2.1a)$$

$$g_{k,l} \;\; = -q\, g_{k,n}\, g_{n,l} - q^{-1}\, g_{n,l}\, g_{k,n} \qquad 1 \le k < n < l \le N \qquad (2.1b)$$

$$g_{l,k} \;\; = q^{-4}\, g_{k,l} \qquad 1 \le k < l \le N. \qquad (2.1c)$$

Note that these operators are non-local. Interchanging $q$ and $q^{-1}$ one can define another set of such operators, but they have the same expectation values. For $q = 1$ and the representation (1.3) the two-point scalar operator reduces to the $SU(2)$ scalar operator

$$g_{k,l}^{(q=1)} = -\frac{1}{2}\vec{\sigma}_k\vec{\sigma}_l\ . \qquad (2.2)$$

For $q = \exp(\frac{i\pi}{m+1})$ a root of unity quotients of the Temperley–Lieb algebra can be used to relate the general hamiltonian (1.1) to other quantum spin chains. Specifically, for $q = \exp(\frac{i\pi}{4})$ the Temperley–Lieb algebra can be represented such that $H$ becomes the hamiltonian of the Ising model with $L = \frac{N}{2}$ sites. With this representation the two-point scalar operators can be expressed as products of fermionic operators and their ground state expectation values can be computed explicitly [2]:

$$\langle g_{2j,2k}\rangle \quad = 0 \qquad (2.3a)$$

$$\langle g_{2j-1,2k-1}\rangle = 0 \qquad (2.3b)$$

$$\langle g_{2j,2k-1}\rangle \quad = -\frac{2\sqrt{2}}{2L+1}\sum_{n=0}^{L-1} \sin\Big(\pi\frac{2n+1}{2L+1}j\Big)\, \cos\Big(\pi\frac{2n+1}{2L+1}(k-\frac{1}{2})\Big) \qquad (2.3c)$$

$$\langle g_{2j-1,2k}\rangle \quad = \frac{2\sqrt{2}}{2L+1}\sum_{n=0}^{L-1} \sin\Big(\pi\frac{2n+1}{2L+1}k\Big)\, \cos\Big(\pi\frac{2n+1}{2L+1}(j-\frac{1}{2})\Big)\ . \qquad (2.3d)$$

These expressions reduce in the limit $L \to \infty$ to

$$\langle g_{k,l}\rangle = \frac{\sqrt{2}}{\pi}\,(\frac{1}{l-k} - \frac{1}{l+k}) \qquad k \text{ even, } l \text{ odd} \qquad (2.4a)$$

$$\langle g_{k,l}\rangle = \frac{\sqrt{2}}{\pi}\,(\frac{1}{l-k} + \frac{1}{l+k}) \qquad k \text{ odd, } l \text{ even}\ . \qquad (2.4b)$$

We will use these results to estimate errors of the numerical computations.

For $q = \exp(\frac{i\pi}{6})$ there exists a different quotient of the Temperley–Lieb algebra such that the hamiltonian $H$ turns into the one of a self-dual three state Potts quantum chain with $\frac{N}{2}$ sites and free boundary conditions [2]. Using this representation the operators $g_{k,l}$ can be rewritten in terms of two different local parafermionic operators, which correspond to respectively even or odd sites of the original quantum chain (1.1). The hamiltonian and expressions for the $g_{k,l}$ are explicitly given in [2]. The parafermionic operators are already known from [17]. However, an analytic solution of the correlation functions using the parafermions has not been achieved.

## 3. Two representations of the Temperley–Lieb algebra

For the numerical computations we use two different graphical representations of the Temperley–Lieb algebra. We describe them in the following and explain how they can be employed to calculate correlation functions.

### *3.1. Path representation*

First let $q$ be generic. Then one can define an action of the Temperley–Lieb algebra on a vector space $\mathcal{S}$ with an orthonormal basis of vectors $v_k$ that are labelled by $N+1$ numbers $k=(k_0,k_1,\ldots,k_N)$, subject to the conditions

$$\begin{aligned} & k_i \geq 0 \\ & k_0 = k_N \ = \ 0 \\ & k_i = k_{i-1} \pm 1/2 \qquad\qquad i=1,\ldots,N \ . \end{aligned} \tag{3.1}$$

On this vector space a generator $e_i$ of $T_N(q)$ can be represented by [11, 16]

$$e_i\ v_k = \delta_{k_{i-1},k_{i+1}} \sum_{k'_i=k_{i-1}\pm\frac{1}{2}} [2k_i+1]_q^{\ 1/2}\ [2k'_i+1]_q^{\ 1/2}\ [2k_{i+1}+1]_q^{\ -1}\ v_{k'} \tag{3.2}$$

with

$$k' = (k_1,\ldots,k_{i-1},k'_i,k_{i+1},\ldots,k_N)\ \ . \tag{3.3}$$

Here we use the definition of the $q$-number

$$[x]_q = \frac{q^x-q^{-x}}{q-q^{-1}}\ .$$

Defining

$$\Gamma_j^{(N)} = \left(\begin{array}{c} N \\ N/2-j \end{array}\right) - \left(\begin{array}{c} N \\ N/2+j+1 \end{array}\right)\ , \tag{3.4}$$

$\mathcal{S}$ has the dimension $\Gamma_0^{(N)}$ [1].

The vectors $v_k$ can be interpreted as paths of a Bratteli diagram [1, 16]. This diagram describes the fusion process of the $N$ spin-$\frac{1}{2}$ representations of $U_qSU(2)$ attached to each site of the spin chain. For the Bratteli diagram $k_i$ gives the spin of an irreducible representation of $U_qSU(2)$ that appears when one decomposes the tensor product of the spin-$k_{i-1}$ multiplet attached to the first $i-1$ sites with the doublet of site $i$. Thus $k_N$ gives the total spin, and the condition $k_N=0$ selects the $U_qSU(2)$ scalars. In this context the $U_qSU(2)$ symmetric generators $e_i$ (1.3) are found to act on the path space according to (3.2).

Next we consider the case $q = \exp(\frac{i\pi}{m+1}) = \exp(\frac{i\pi r}{s})$ with $r$ and $s$ coprime integers. Since $[s]_q = 0$, in this case a basis of the path representation space of $T_N(q)$ is given by the vectors $v_k$ with

$$k_i \leq \frac{s}{2} - 1. \tag{3.5}$$

We denote this vector space as $\mathcal{S}_q \subset \mathcal{S}$. This reduction is reflected by the $U_qSU(2)$ representations for $q$ a root of unity [1]. In this case indecomposable but reducible representations appear in the decomposition of the spin configuration space. Only the spin zero representations whose paths in the Bratteli diagram are restricted according to (3.5) remain as irreducible representations. Their number, i.e. the dimension of $\mathcal{S}_q$, is given by [1]

$$\Omega_m^{(N)} = \Gamma_0^{(N)} - \Gamma_{(m+1)-1}^{(N)} + \Gamma_{m+1}^{(N)} - \Gamma_{2(m+1)-1}^{(N)} + \Gamma_{2(m+1)}^{(N)} - \Gamma_{3(m+1)-1}^{(N)} + \cdots \tag{3.6}$$

and thus depends on $q = \exp(\frac{i\pi}{m+1})$. Since the ground state of the hamiltonian (1.1) is known to be non-degenerate, we can restrict the representation space for the numerical calculation of the correlation functions to $\mathcal{S}$ respectively $\mathcal{S}_q$. The action of hamiltonian (1.1) and two-point scalar operator (2.1*a*-*c*) on the path space follow from the definition (3.2).

### *3.2. Representation on boundary diagrams*

In [2] a different diagrammatic approach to the calculation of correlation functions $\langle g_{k,l} \rangle$ has been proposed, that uses the regular representation of the Temperley–Lieb algebra on boundary diagrams [16]. Boundary diagrams are given by $N$ non-crossing lines that connect $N$ upper and $N$ lower points such that each point is connected to one other point. For the generators we have the definition:

$e_i =$ 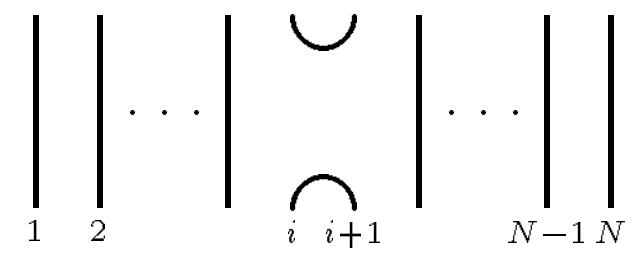

. (3.7)

The diagrams corresponding to words of the algebra are obtained as follows. For the composition of two words the corresponding diagrams are stacked on top of each other and the lines ending in the lower points of the first diagram are joined with the lines ending in the corresponding upper points of the second diagram. Any closed line appearing in this process is discarded from the diagram and replaced by a factor $q + q^{-1}$. This reflects the Temperley–Lieb relation (1.2*a*). The other Temperley–Lieb relations (1.2*b*) and (1.2*c*) can easily be verified by drawing the corresponding diagrams. In this way elements of the Temperley–Lieb algebra can be represented faithfully as linear combinations of boundary diagrams [16].

Let us restrict this regular representation of $T_N(q)$ to the one-sided ideal $\mathcal{I}$ generated by the word $e_1 e_3 \ldots e_{N-1}$. We have

$$e_1 e_3 \ldots e_{N-1} = \qquad\qquad (3.8)$$

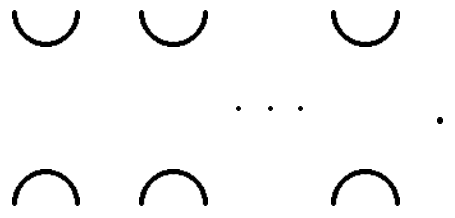

The ideal $\mathcal{I}$ is known to represent the space of $U_qSU(2)$ scalars for generic $q$ [16]. Hence the ground state can be represented by a vector in $\mathcal{I}$. The dimension of $\mathcal{I}$ (as of $\mathcal{S}$) is given by $\Gamma_0^{(N)}$. The ideal is realized by all diagrams with disconnected upper and lower parts where the lower part is the one in (3.8) and the upper part has any configuration of $\frac{N}{2}$ non-intersecting lines between the $N$ upper points.

To calculate expectation values one needs a scalar product on $\mathcal{I}$. Define a transposed diagram as the upside-down reflected diagram. With this definition the scalar product $\langle s_1 | s_2 \rangle$ of two arbitrary elements $s_1$ and $s_2$ of $\mathcal{I}$ is defined [2]:

$$s_1{}^T\, s_2 \;=\; \langle s_1 | s_2 \rangle \qquad\qquad (3.9)$$

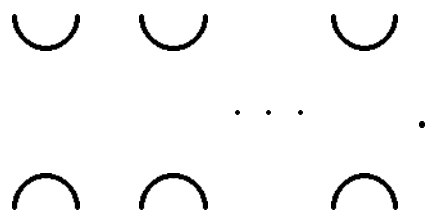

That is, the scalar product is given by the factor that one obtains removing any loop from the diagrams which represent $s_1{}^T\, s_2$.

With the definition of boundary diagrams given above and the scalar product (3.9), the expectation values $\langle v | g_{k,l} | v \rangle$ of the correlation operators can be computed by multiplying diagrams, that encode the Temperley–Lieb relations. This way it can easily be seen that $\langle v | g_{k,l} | v \rangle$ is real for any vector $v$ in $\mathcal{I}$ .

For $q$ a root of unity quotients of the Temperley–Lieb algebra have been given in [15]. The calculation of the ground state average of $g_{k,l}$ through evaluation of the corresponding diagrams according to (3.9) is independent of the quotienting, since it depends on the Temperley–Lieb relations (1.2*a*-*c*) only. In principle the quotients can be used to eliminate certain vectors from $\mathcal{I}$. However, this is not useful for the numerical calculations, since the elimination procedure would have to be carried out after each application of a generator to a vector of $\mathcal{I}$.

## 4. Computational method

Let us describe the numerical calculation of the ground state expectation values $\langle g_{k,l} \rangle$ of the two-point operators (2.1*a*-*c*). We employ the representations of section 3 so that we do not have to refer to the spin configuration space. In this article we present the cases $q = \exp(\frac{i\pi}{m+1})$ with $m = 3$, $m = 5$, and $m = \frac{2}{3}$, but the computation can be performed analogously for any $m$.

First we generate a basis of the configuration space $\mathcal{I}$ (boundary diagrams) or $\mathcal{S}_q$ (path representation) and the matrix of the hamilton operator in the respective basis. In general the path representation has the advantage of a smaller representation space and an orthonormal basis of this space. Note that the matrices are not hermitian (although the eigenvalues of $H$ are real) so that we cannot use the Lanczös algorithm. We employ a power method to find the ground state: repeated application of the matrix to an appropriate start vector projects onto the eigenstate with the eigenvalue of highest absolute value. It requires about 2000 iterations and gives the eigenenergy with a relative error of less than $10^{-11}$ ($10^{-18}$ for small $N$). We have only applied this method in the cases where $H$ is represented by a real matrix. The complex case requires more storage capacity and multiplications. For the representation on boundary diagrams the matrix realizing $H$ is real independent of the value of $q$. With the path representation we have a real hamiltonian for integer $m$. For the non-unitary model given by $m = \frac{2}{3}$ the path representation gives a complex hamiltonian so that we use only the boundary diagrams. Note that for $m = \frac{2}{3}$ the norm of the ground state is negative according to (3.9).

For $m = 3$ and $m = 5$ the ground state energy was checked against Bethe ansatz calculations [13, 18]. Our results (obtained up to $N$=28) are identical to the results listed in [13]. Also, using the formulas given in [12, 13], extrapolation from the finite size energies gives the correct values (1.4) of the conformal charge. We find $c = 0.50(1)$ for $m = 3$, $c = 0.79(1)$ for $m = 5$, and $c = -4.4(1)$ for $m = \frac{2}{3}$ using the BST extrapolation algorithm [19, 20].

We find that the ground state has positive parity, i.e. the following property: The contribution of a specific vector $v_k$ from $\mathcal{S}_q$ (a word from $\mathcal{I}$) to the ground state is the same as the one of the vector labelled by $k' = (k_N, k_{N-1}, \ldots, k_1)$ (the word with left and right ends exchanged).

For the computation of the correlation functions we make explicit use of the recursion in (2.1*a*-*c*). We also use that the expectation values $\langle v|g_{k,l}|v\rangle$ of the operators $g_{k,l}$ have the symmetry

$$\langle v|g_{k,l}|v\rangle = \langle v|g_{N+1-l,N+1-k}|v\rangle \tag{4.1}$$

for any vector $v$ with positive or negative parity. This can most easily be seen employing the boundary diagrams as a regular representation and using the reality of $\langle v|g_{k,l}|v\rangle$.

On a workstation we are able to calculate the correlations for chains of up to $N = 24$ sites with the representation using boundary diagrams (independent of $q$). In this case the representation space has dimension $\Gamma_0^{(24)} = 208012$. The limit is set by the time required for the computation of the scalar products (3.9). For the path space representation the storage capacity limits the calculations. In this case the dimension of the representation space depends on $m$. We can handle $N = 40$ sites for $m = 3$ ($\dim \mathcal{S}_q = \Omega^{(40)}_{m=3} = 524288$) and $N = 28$ sites for $m = 5$ ($\Omega^{(28)}_{m=5} = 797162$).

From the known results in the Ising case ($m = 3$) we can determine the numerical errors of the computed correlations. The results are exact to 14 digits for $N = 14$ and to 11 digits for $N = 28$. The two approaches give the same values up to 12 digits ($N = 24$).

We use the BST algorithm to extrapolate the correlation functions for finite $N$ to the values of the semi-infinite chain ($N \to \infty$), i.e. we extrapolate the values $\langle g_{k,l} \rangle$ for different $N$ with $k$ and $l$ fixed. With the given precision of our numerical calculation we need at least the results of 6 different chain lengths to obtain reliable results (4 digits) with the BST extrapolation. This can be seen from the convergence behaviour of the BST algorithm. In the case $m = 3$ we can also check the extrapolated results using (2.4*a*-*b*) and find the same accuracy. Hence for the semi-infinite chain we can give $\langle g_{k,l} \rangle$ with $k, l \leq 18$ ($m = 5$) and $k, l \leq 14$ ($m = 2/3$). These values are listed in the appendix.

## 5. Results

Here we discuss the correlation functions, i.e. the ground state expectation values $\langle g_{k,l} \rangle$, for the three cases $q = \exp(\frac{\mathrm{i}\pi}{m+1})$ with $m = 3, 5, \frac{2}{3}$. We consider the extrapolated results for the semi-infinite chain. For $q$ on the unit circle the model is massless and thus we expect long-range order with power-law decay of the correlation functions. In the bulk limit (without boundary influence) we expect

$$\langle g_{k,l} \rangle \propto \frac{1}{(l-k)^{2x}} \tag{5.1}$$

with the standard definition of a bulk critical exponent $x$. From this and the relation $g_{k,l} = q^4 g_{l,k}$ for $k < l$ we find that

$$x = \frac{m-1}{m+1} \; . \tag{5.2}$$

The continuum limit of the correlation functions of the semi-infinite chain corresponds to two-point functions of a conformal field theory in the half plane with central charge given by (1.4). In these minimal models the highest weights are found to be [21]

$$h_{p,q}(m) = \frac{[(m+1)p - mq]^2 - 1}{4m(m+1)}, \tag{5.3}$$

for $q$ and $p$ integers. Comparing (5.2) with (5.3) we are forced to take

$$x = h_{1,3}(m) \quad . \tag{5.4}$$

This implies that in the continuum limit the correlation function $\langle g_{k,l} \rangle$ is realized by two-point functions of linear combinations of conformal fields with dimension $(h_{1,3}, 0)$ or $(0, h_{1,3})$. For $m = 3$ the analytic result is known and one actually finds Majorana fields with spin $\frac{1}{2}$ [2, 22]. Nonetheless we describe the analysis of the numerical results for this simple case, in order to test methods of finding information about the continuum limit of the correlation functions. By means of the numerical investigations we try to confirm (5.4) for other values of $m$.

### *5.1. $m = 3$ (Ising model)*

The ground state averages $\langle g_{k,l} \rangle$ show a different behaviour for $k$ and $l$ odd or even respectively. The values $\langle g_{k,l} \rangle$ are positive for $l - k$ odd (with a different behaviour for $k$ odd or even) and zero otherwise. Thus for the continuum limit of the correlation functions one has to make a distinction between even and odd sites. From the analytic result for the semi-infinite chain (2.4) we find that for both correlation functions

$$\langle g_{k,l} \rangle = \frac{1}{(l-k)} F(k/l) \; . \tag{5.5}$$

The bulk limit is given by $k \to \infty$ with $l - k$ fixed. Thus $\frac{k}{l} \to 1$ and we have $x = \frac{1}{2}$ from (5.1), corresponding to the Majorana fields. Note that the form (5.5) of the correlation functions is valid for arbitrary $k$ and $l$. The scaling function $F$ describing the influence of the boundary depends only on the ratio $\frac{k}{l}$. This can be expected of the continuum limit from general scaling arguments. In general the surface critical exponent $x_s$ is given by [23]

$$F(k/l) \propto (k/l)^{x_s - x} \;\; \text{for} \;\; \frac{k}{l} \to 0 \; . \tag{5.6}$$

From (2.4) one has $x_s = \frac{1}{2}$ for the correlation function $\langle g_{2j-1,2k} \rangle$ and $x_s = \frac{3}{2}$ for $\langle g_{2j,2k-1} \rangle$.

Next we examine the numerical results extrapolated to the semi-infinite chain. The scaling behaviour of $F$ can be seen from figure 1($a$). In this figure we plot the contour lines $\langle g_{k,l} \rangle \, (k-l)^{2x}$ for $k$ even and $l$ odd. The straight contour lines passing through the origin imply a dependence of $F$ on $\frac{k}{l}$ only. The other correlation function with $k$ odd and $l$ even shows the same behaviour. Assuming that $F$ is a function of $\frac{k}{l}$ we can calculate the bulk critical exponent $x$ from the numerical data. Using extrapolated values for the functions $\langle g_{2j,2k-1} \rangle$ and $\langle g_{2j-1,2k} \rangle$ on the semi-infinite chain, we find $x = 0.5000(1)$ in both cases. However, the surface exponent $x_s$ can only be estimated with an accuracy of about 15% from the numerical data (extrapolated from chains of up to 40 sites), because the limit in (5.6) can not be reached with the available values $k$ and $l$.

### *5.2. $m = 5$ (three state Potts model)*

Next we consider the case $q = \exp(\frac{i\pi}{6})$, which corresponds to a three state Potts model as described in section 2. The computed $\langle g_{k,l} \rangle$ are positive for $l - k$ odd and negative otherwise. We find that we have to distinguish four non-zero correlation functions. That is, as in the case $m = 3$, the correlation functions described by $|\langle g_{k,l} \rangle|$ are different for $k$ and $l$ odd or even. The appearance of different functions for odd or even sites can be expected from the expressions for the correlation operators in terms of parafermions that arise from a quotient representation of the Temperley–Lieb algebra for $q = \exp(\frac{i\pi}{6})$ (section 2).

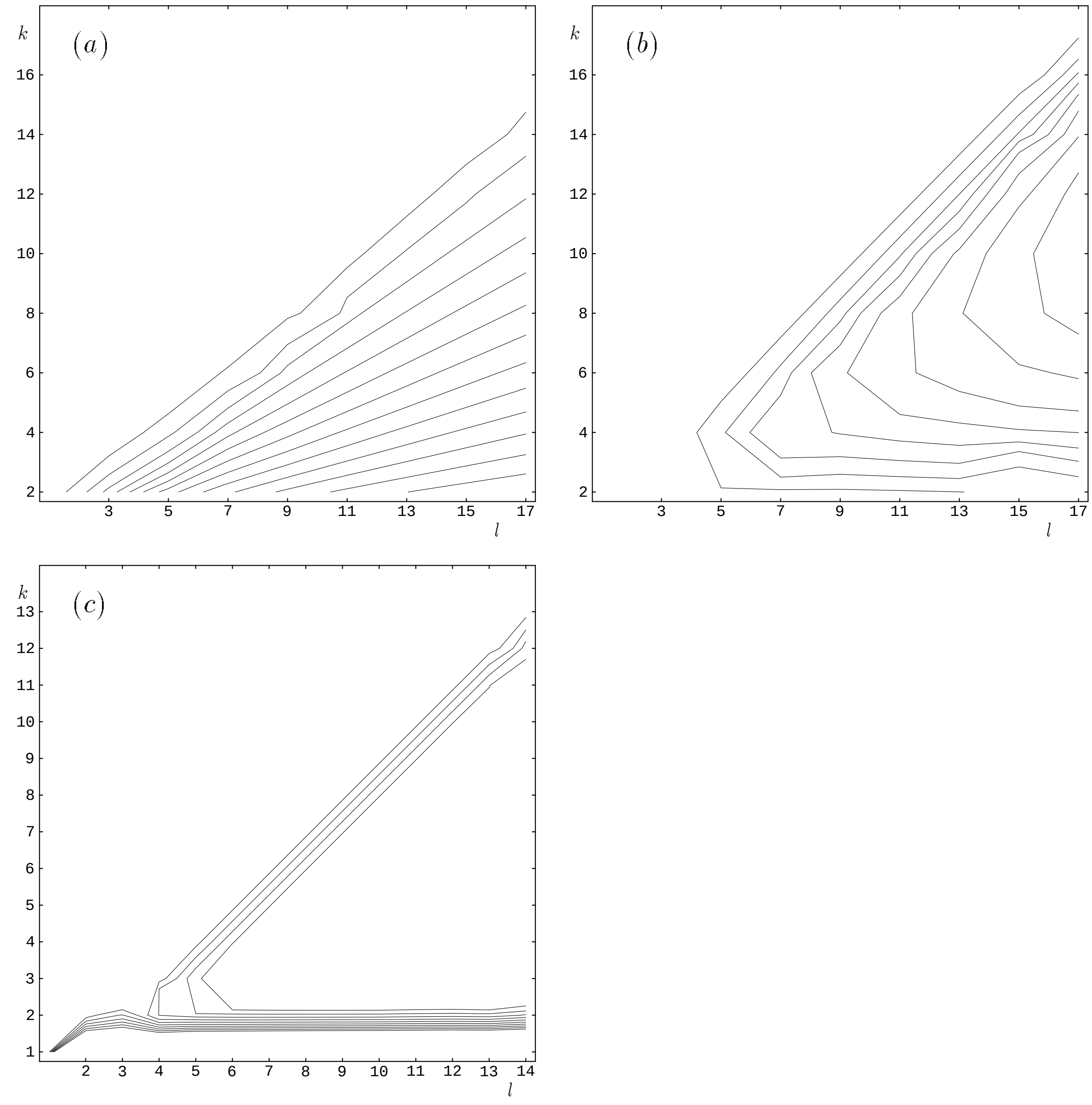

**Figure 1.** Contour lines of $\langle g_{k,l}\rangle(l-k)^{2x}$ for the semi-infinite chain. $(a)$ shows the data for $q=\exp(\frac{\mathrm{i}\pi}{4})$ where $x=\frac{1}{2}$, $(b)$ the data for $q=\exp(\frac{\mathrm{i}\pi}{6})$ with $x=\frac{2}{3}$, and $(c)$ the data for $q=\exp(\frac{\mathrm{i}\pi 3}{5})$ where $x=-\frac{1}{5}$. For $(a)$ and $(b)$ we use the values for $k$ even and $l$ odd with $k<l\leq 18$, for $(c)$ we use all values with $k<l\leq 14$. The spacing between contour lines is 0.025 $(a)$, 0.05 $(b)$ and 0.01 $(c)$.

From (5.2) we expect the bulk critical exponent $x = \frac{2}{3}$. In figure 1($b$) we give the contour lines of $F(k,l) = g_{k,l}\,(l-k)^{4/3}$ for $k$ even and $l$ odd. This plot is representative of the following: For the values $k, l \leq 18$ (which we can determine the correlations of the semi-infinite chain for) none of the four scaling functions $F$ can be described by a function of $\frac{k}{l}$. Furthermore we find that there is no value of $x$ such that $F$ is a function of $\frac{k}{l}$ only. We conclude, that the range of $k, l$ is too small to be in the scaling regime. In summary, the critical exponents $x$ and $x_s$ cannot be computed in the case $m = 5$. This does not result from inaccuracies of computation and extrapolation of the correlations, but from the fact that in this case the chain length is not sufficient to give the continuum limit.

### *5.3. $m = \frac{2}{3}$ (Lee–Yang edge singularity)*

Here $q = \exp(\frac{i\pi 3}{5})$ and the hamiltonian $H$ is related to a conformal field theory with central charge $c = -\frac{22}{5}$. This non-unitary minimal model describes the Lee–Yang singularity of an Ising model in an imaginary longitudinal magnetic field [24]. Besides the identity it has only one primary operator with conformal dimension $h = -\frac{1}{5}$. This is the dimension we expect as the bulk exponent from equation (5.2).

The correlations $\langle g_{l,m} \rangle$ show a behaviour different from the minimal models $m = 3$ and $m = 5$. We find only one (positive) correlation function independent of $k, l$ even or odd. In figure 1($c$) we plot contour lines for all values of $k, l \leq 14$. The scaling function $F(k,l) = \langle g_{k,l} \rangle (l-k)^{-2/5}$ becomes constant for $k \gg 1$ and $l - k \gg 1$. Thus we expect $F(k/l) = \text{const.}$ in the continuum limit yielding $x_s = x$.

On the lattice there is an influence of the boundary, which vanishes quickly with $k > 2$. With the correlation function of the semi-infinite chain one obtains the bulk exponent from (5.1), taking $k \geq 3$ and $l - k \geq 4$ and assuming a constant $F$. This way we find $x = -0.20(1)$ verifying (5.4).

## 6. Conclusion

We have performed numerical investigations on the $U_qSU(2)$ scalar two-point operator for the $U_qSU(2)$ symmetric $XXZ$ quantum chain at roots of unity. The properties of the correlation functions on the chain depend strongly on the value of $q$ as we show for three different cases. For $q = \exp(\frac{i\pi}{4})$ one can find the (known) bulk exponent of the two correlation functions $\langle g_{2j,2k-1}\rangle$ and $\langle g_{2j-1,2k}\rangle$ from small chains. For the non-unitary model given by $q = \exp(\frac{i\pi 3}{5})$ it is possible to extract information about the continuum limit and critical exponents from chains with up to 24 sites. We have one correlation function and find the bulk critical exponent $-\frac{1}{5}$ confirming $x = h_{1,3} = \frac{m-1}{m+1}$. Thus the continuum limit of the correlations involves operators with spin $h_{1,3}$ as in the case $q = \exp(\frac{i\pi}{4})$, where this is already known from an analytical solution [2]. For the surface exponent we obtain $x_s = -\frac{1}{5}$ as well. For $q = \exp(\frac{i\pi}{6})$, where the $XXZ$ chain corresponds to a three state Potts model, we find four different correlation functions $\langle g_{k,l}\rangle$ depending on $k, l$ even or odd. In this case it is impossible to find the continuum limit from the lattice data (up to 28 sites). Thus the operators corresponding to the parafermions on the spin chain can not be found in this way. We have performed analogous computations for other integer $m \geq 4$. We do not present these results here, but want to remark that lattice effects dominate the correlations in these cases as well. Thus in general it remains an open problem to find the conformal operators corresponding to the $U_qSU(2)$ invariant correlation operators for the quantum spin chain.

## Acknowledgments

We would like to thank Vladimir Rittenberg and Fabian Eßler for many helpful discussions.

## Appendix

We list the ground state expectation values $\langle g_{k,l}\rangle$ of the two-point scalar operators (2.1) extrapolated to the semi-infinite chain for the cases $q = \exp(\frac{i\pi 3}{5})$ corresponding to $c = -\frac{22}{5}$ (table 1) and $q = \exp(\frac{i\pi}{6})$ corresponding to $c = \frac{4}{5}$ (table 2). Estimating the error as explained in section 4 these results are reliable up to four digits as given in the tables. We do not give the correlations in the case $q = \exp(\frac{i\pi}{4})$. They can be calculated from (2.4).

**Table 1.** Correlation functions $\langle g_{k,l}\rangle$ of the semi-infinite chain with $k < l \leq 14$ for $m = \frac{2}{3}$ (Lee–Yang edge singularity). The values given are extrapolated from chains with up to 24 sites.

| $k \setminus l$ | 2 | 3 | 4 | 5 | 6 | 7 | 8 | 9 | 10 | 11 | 12 | 13 | 14 |
|---|---|---|---|---|---|---|---|---|---|---|---|---|---|
| 1 | 1.81 | 2.392 | 2.788 | 3.108 | 3.384 | 3.629 | 3.85 | 4.054 | 4.244 | 4.418 | 4.578 | 4.746 | 4.879 |
| 2 | | 2.026 | 2.714 | 3.2 | 3.592 | 3.928 | 4.225 | 4.494 | 4.74 | 4.965 | 5.177 | 5.381 | 5.551 |
| 3 | | | 2.041 | 2.737 | 3.23 | 3.627 | 3.968 | 4.269 | 4.541 | 4.788 | 5.015 | 5.236 | 5.42 |
| 4 | | | | 2.042 | 2.739 | 3.232 | 3.63 | 3.971 | 4.272 | 4.543 | 4.793 | 5.024 | 5.224 |
| 5 | | | | | 2.043 | 2.739 | 3.232 | 3.63 | 3.971 | 4.272 | 4.544 | 4.793 | 5.012 |
| 6 | | | | | | 2.043 | 2.739 | 3.232 | 3.631 | 3.971 | 4.272 | 4.544 | 4.784 |
| 7 | | | | | | | 2.043 | 2.739 | 3.232 | 3.63 | 3.971 | 4.272 | 4.544 |
| 8 | | | | | | | | 2.043 | 2.739 | 3.232 | 3.63 | 3.97 | 4.27 |
| 9 | | | | | | | | | 2.043 | 2.739 | 3.232 | 3.63 | 3.971 |
| 10 | | | | | | | | | | 2.043 | 2.739 | 3.232 | 3.63 |
| 11 | | | | | | | | | | | 2.043 | 2.739 | 3.231 |
| 12 | | | | | | | | | | | | 2.043 | 2.734 |
| 13 | | | | | | | | | | | | | 2.043 |

**Table 2.** Correlation functions $\langle g_{k,l} \rangle$ of the semi-infinite chain with $k < l \leq 18$ for $m = 5$ (three state Potts model). The values given are extrapolated from chains with up to 28 sites.

| $k \setminus l$ | 2 | 3 | 4 | 5 | 6 | 7 | 8 | 9 | 10 |
|---|---|---|---|---|---|---|---|---|---|
| 1 | 0.9907 | −0.1764 | 0.3546 | −0.1064 | 0.2139 | −0.07631 | 0.1514 | −0.05946 | 0.1162 |
| 2 | | 0.5136 | −0.1458 | 0.1247 | −0.08158 | 0.06304 | −0.05589 | 0.04036 | −0.04212 |
| 3 | | | 0.8489 | −0.1647 | 0.3074 | −0.1004 | 0.1918 | −0.0733 | 0.1397 |
| 4 | | | | 0.584 | −0.1542 | 0.156 | −0.08962 | 0.0822 | −0.06306 |
| 5 | | | | | 0.8051 | −0.162 | 0.287 | −0.09839 | 0.1798 |
| 6 | | | | | | 0.6139 | −0.1566 | 0.1714 | −0.09219 |
| 7 | | | | | | | 0.783 | −0.1609 | 0.2755 |
| 8 | | | | | | | | 0.6309 | −0.1575 |
| 9 | | | | | | | | | 0.7694 |

| $k \setminus l$ | 11 | 12 | 13 | 14 | 15 | 16 | 17 | 18 |
|---|---|---|---|---|---|---|---|---|
| 1 | −0.04866 | 0.09376 | −0.04115 | 0.07825 | −0.03561 | 0.06696 | −0.03138 | 0.05834 |
| 2 | 0.02907 | −0.03358 | 0.02246 | −0.02778 | 0.01819 | −0.02361 | 0.01522 | −0.02047 |
| 3 | −0.05799 | 0.1096 | −0.04806 | 0.09002 | −0.04106 | 0.07617 | −0.03586 | 0.06595 |
| 4 | 0.05365 | −0.04847 | 0.03902 | −0.03923 | 0.03028 | −0.03285 | 0.02456 | −0.02819 |
| 5 | −0.0719 | 0.1319 | −0.05707 | 0.1044 | −0.04748 | 0.08643 | −0.04073 | 0.07369 |
| 6 | 0.0925 | −0.06561 | 0.06122 | −0.05092 | 0.0449 | −0.04155 | 0.03504 | −0.03504 |
| 7 | −0.09741 | 0.1722 | −0.07113 | 0.1267 | −0.05649 | 0.1006 | −0.04706 | 0.08364 |
| 8 | 0.1808 | −0.09336 | 0.09908 | −0.06684 | 0.06623 | −0.05215 | 0.04891 | −0.04276 |
| 9 | −0.1603 | 0.2679 | −0.09687 | 0.167 | −0.07072 | 0.1228 | −0.05611 | 0.0977 |
| 10 | 0.642 | −0.158 | 0.1872 | −0.09399 | 0.1037 | −0.06753 | 0.06984 | −0.05287 |
| 11 | | 0.7602 | −0.16 | 0.2626 | −0.09654 | 0.1632 | −0.07037 | 0.1199 |
| 12 | | | 0.6499 | −0.1583 | 0.1918 | −0.09437 | 0.1071 | −0.06796 |
| 13 | | | | 0.7534 | −0.1598 | 0.2586 | −0.09632 | 0.1603 |
| 14 | | | | | 0.6557 | −0.1585 | 0.1954 | −0.09462 |
| 15 | | | | | | 0.7482 | −0.1597 | 0.2555 |
| 16 | | | | | | | 0.6604 | −0.1587 |
| 17 | | | | | | | | 0.7441 |